\documentclass[pre,aps,preprint,showpacs,preprintnumbers,amsmath,amssymb,secnumarabic,eqsecnum]{revtex4}

\usepackage{amsmath}
\usepackage{amssymb}
\usepackage{graphicx}
\usepackage{dcolumn}
\usepackage{bm}

\newcommand{\beq}{\begin{equation}}
\newcommand{\eeq}{\end{equation}}
\newcommand{\beqa}{\begin{eqnarray}}
\newcommand{\eeqa}{\end{eqnarray}}
\newcommand{\vc}[1]{\mbox{\boldmath $#1$}}

\newcommand{\vol}[1]{{\bf #1}}

\begin{document}


\title{Self-propulsion of a spherical electric or magnetic microbot in a polar viscous fluid}

\author{B. U. Felderhof}

 \email{ufelder@physik.rwth-aachen.de}
\affiliation{Institut f\"ur Theorie der Statistischen Physik\\ RWTH Aachen University\\
Templergraben 55\\52056 Aachen\\ Germany\\
}%

\date{\today}

\begin{abstract}
The self-propulsion of a sphere immersed in a polar liquid or ferrofluid is studied on the basis of ferrohydrodynamics. In the electrical case an oscillating charge density located inside the sphere generates an electrical field which polarizes the fluid. The lag of polarization with respect to the electrical field due to relaxation generates a time-independent electrical torque density acting on the fluid causing it to move. The resulting propulsion velocity of the sphere is calculated in perturbation theory to second order in powers of the charge density.
\end{abstract}

\maketitle

\section{\label{I}Introduction}

A planar sheet can propel itself through a polar liquid or a ferrofluid by the generation of an oscillating electric or magnetic polarization in the fluid \cite{1}. The mechanism is based on the nonlinear coupling of polarization and field, which leads to a torque density acting on the fluid, causing it to move. The same coupling allows pumping of a neutral electrically polar liquid like water by the application of a running electric wave \cite{2}, and pumping of a ferrofluid by the application of a running magnetic wave \cite{3},\cite{4}. In the following we study electric or magnetic self-propulsion of a spherical microbot. The planar sheet provides a simple model allowing straightforward mathematical analysis, but a sphere resembles more closely a compact object which would be used experimentally and needs to be analyzed separately.

For definiteness we discuss only the electrical case in detail, but a quite analogous derivation holds in the magnetic case. In the electrical case we assume that the motion is caused by an oscillating charge density located inside a rigid spherical surface, generating an electrical field which polarizes the surrounding fluid. For simplicity we assume that the electrical field can be decomposed into dipole and quadrupole contributions. The sphere is caused to move by interference of the dipole and quadrupole fields which leads to a time-independent electrical azimuthal torque density acting on the fluid. The torque density creates a steady vortex ring surrounding the sphere and propelling it. In Fig. 1 we draw a schematic picture of the effect.

In analogy to the theory of ferrohydrodynamic pumping \cite{5} we calculate the propulsion velocity of the sphere in perturbation theory to second order in powers of the amplitude of the exciting charge density. The perturbation calculation has the advantage of simplicity. It leads to an explicit expression for the propulsion velocity, and hence allows insight into its dependence on the system parameters. To the order considered, the nonlinear convective terms in the equations of motion can be neglected. We have shown in the case of ferrohydrodynamic pumping \cite{5} and self-propulsion of a planar microbot \cite{1} that the theory can be extended to higher order, but the second order perturbation theory turned out to be quite sufficient from a numerical point of view. Therefore we limit the calculation to second order perturbation theory in the present case. At the surface of the sphere the flow velocity is required to satisfy a no-slip boundary condition, but a mixed slip-stick boundary condition could also be considered.

The spherical geometry is preferable to the planar one in experimental realization and in numerical simulation. It would be fascinating to construct a microbot without moving parts of the type considered here. If realized, it may offer interesting technical application, for example in the transport of drugs through a fluid. The physical situation under consideration provides a remarkable example of the coupling of translational and rotational degrees of freedom in a fluid \cite{6}-\cite{9}. The spherical geometry poses a challenging problem and the mathematical solution has an interest of its own.

The spatial shift of the spherical microbot during a period of the field is related to the concept of geometric phase or holonomy \cite{9A}. Examples of holonomy are Foucault's pendulum, the four bar linkage studied by Yang and Krishnaprasad \cite{9B}, and Berry's phase in quantum mechanics \cite{9C}.

The self-propulsion studied here is more closely related to swimming than to phoresis in an applied field. In electrokinetic phenomena in electrolyte solutions the effects are linear in the applied electric field to lowest order \cite{9D}. In the present case the propulsion velocity is quadratic in the self-generated field, as in swimming at low Reynolds number \cite{9E},\cite{9F}, where to lowest order the speed is quadratic in the amplitude of surface distortion. We study a single active particle, but in principle two or more with hydrodynamic interactions may be considered \cite{9G}.

\section{\label{II}Equations of motion}

We consider a sphere of radius $a$ immersed in an incompressible polar viscous fluid with shear viscosity $\eta$, vortex viscosity $\zeta$, and spin viscosity $\eta'$. The fluid can be either electrically or magnetically polar. For definiteness we shall use language appropriate to an electrically polar liquid. With minor changes the same equations apply in the case of a magnetic ferrofluid.

Due to incompressibility of the fluid the divergence of the flow velocity $\vc{v}(\vc{r},t)$ vanishes, $\nabla\cdot\vc{v}=0$. The flow velocity satisfies the momentum
balance equation
\begin{equation}
\label{2.1}\rho\frac{d\vc{v}}{dt}=\nabla\cdot(\vc{\sigma}_{hyd}+\vc{\sigma}_{el}),
\end{equation}
where $d/dt=\partial/\partial t+\vc{v}\cdot\nabla$ is the substantial derivative, $\vc{\sigma}_{hyd}$ is the hydrodynamic stress tensor and
$\vc{\sigma}_{el}$ is the Maxwell stress tensor. The hydrodynamic stress tensor has Cartesian components \cite{10},\cite{11}
\begin{equation}
\label{2.2}\sigma_{hyd,\alpha\beta}=-p\delta_{\alpha\beta}+\eta(\partial_\alpha v_\beta+\partial_\beta v_\alpha)+\zeta\epsilon_{\alpha\beta\gamma}(\nabla\times\vc{v}-2\vc{\omega}_p)_\gamma,
\end{equation}
where $p$ is the pressure, $\eta$ is the shear viscosity, $\zeta$ is the vortex viscosity \cite{8}, and $\vc{\omega}_p$ is the rate of rotation of the polar molecules. In SI units
the Maxwell stress tensor has the form \cite{12}
\begin{equation}
\label{2.3}\vc{\sigma}_{el}=\vc{D}\vc{E}-\frac{\varepsilon_1}{2}E^2\vc{I},
\end{equation}
where $\vc{D}(\vc{r},t)$ is the electric displacement,
$\vc{E}(\vc{r},t)$ is the electrical field, $\varepsilon_1$ is the high-frequency dielectric permeability of the fluid, $E^2=\vc{E}\cdot\vc{E}$, and $\vc{I}$ is the unit tensor. The fields are related by
\begin{equation}
\label{2.4}\vc{D}=\varepsilon_1(\vc{E}+\vc{P}),
\end{equation}
where $\vc{P}(\vc{r},t)$ is the polarization due to permanent dipole moments of the fluid molecules. The fields satisfy Maxwell's equations
of electrostatics
\begin{equation}
\label{2.5}\nabla\cdot\vc{D}=\rho_{el},\qquad\nabla\times\vc{E}=0,
\end{equation}
where $\rho_{el}(\vc{r},t)$ is the electrical charge density located inside the sphere. The charge density $\rho_{el}$ acts as a source of the fields, and is assumed to be known.  We use spherical coordinates $(r,\theta,\varphi)$ with the origin located at the center of the sphere.
The charge density is taken to be a superposition of dipole and quadrupole components such that the first order electric field outside the sphere is given by
\begin{equation}
\label{2.6}\vc{E}_1(r,\theta,t)=\sum^2_{l=1}\mu_l(t)\vc{u}_l(r,\theta),\qquad r>a,
\end{equation}
with dipole moment $\mu_1(t)$, quadrupole moment $\mu_2(t)$, and component field
\begin{equation}
\label{2.7}\vc{u}_l(r,\theta)=\bigg(\frac{a}{r}\bigg)^{l+2}\big[(l+1)P_l(\cos\theta)\vc{e}_r+P^1_l(\cos\theta)\vc{e}_\theta\big],
\end{equation}
with Legendre polynomials $P_l$ and associated Legendre functions of the first kind $P_l^1$ in the notation of Edmonds \cite{13}. The electrical field $\vc{E}_1$ can be derived from a scalar potential $\phi_1$ as $\vc{E}_1=-\nabla\phi_1$ by use of the identity
\begin{equation}
\label{2.8}\vc{u}_l(r,\theta)=-a^{l+2}\nabla\Phi^-_l(r,\theta),\qquad\Phi^-_l(r,\theta)=r^{-l-1}P_l(\cos\theta).
\end{equation}
We assume that the multipole moments vary harmonically in time with frequency $\omega$ and can be expressed as
\begin{equation}
\label{2.9}\mu_l(t)=\mu_{lc}\cos\omega t+\mu_{ls}\sin\omega t,\qquad (l=1,2).
\end{equation}
The first order electrical field $\vc{E}_1(\vc{r},t)$ has the character of a running wave. The external multipole moments $\mu_1,\mu_2$ are linear in the exciting charge density $\rho_{el}$, and must be calculated from an electrostatic problem with account of the high-frequency permeability $\varepsilon_1$ and the first order polarization $\vc{P}_1$. The details of the linear electrostatic problem of a sphere immersed in a dielectric medium need not concern us here.

The relaxation of polarization $\vc{P}$ is assumed to be governed by the constitutive equation \cite{12}
\begin{equation}
\label{2.10}\frac{\partial\vc{P}}{\partial t}+\vc{v}\cdot\nabla\vc{P}-\vc{\omega}_p\times\vc{P}=-\gamma[\vc{P}-\vc{P}_{eq}(\vc{E})],
\end{equation}
where $\vc{P}_{eq}(\vc{E})$ is given by the equilibrium equation of state, and the relaxation rate $\gamma$ is the inverse of the relaxation time $\tau$. The rotation rate $\vc{\omega}_p$ is related to the spin $\vc{S}$ per unit mass by $\vc{S}=I\vc{\omega}_p$, where $I$ is an average moment of inertia per unit mass. The equation of motion for the spin per unit mass is taken as
\begin{equation}
\label{2.11}\rho\frac{d\vc{S}}{dt}=2\zeta(\nabla\times\vc{v}-2\vc{\omega}_p)+\varepsilon_1\vc{P}\times\vc{E}+\eta'\nabla^2\vc{\omega}_p,
\end{equation}
where $\eta'$ is the spin viscosity \cite{12}. The first term on the right is the hydrodynamic torque density, and the second term is the electrical torque density. In the situation considered in the following $\nabla\cdot\vc{\omega}_p=0$ due to spatial symmetry, so that there is no need to introduce a bulk spin viscosity \cite{6}.

 We shall neglect the inertial term on the left-hand side in Eqs. (2.1) and (2.11). Then Eq. (2.11) reduces to
 \begin{equation}
\label{2.12}2\zeta(\nabla\times\vc{v}-2\vc{\omega}_p)=-\varepsilon_1\vc{P}\times\vc{E}-\eta'\nabla^2\vc{\omega}_p.
\end{equation}
Substituting this into Eq. (2.2) we find from Eq. (2.1)
\begin{equation}
\label{2.13}\eta\nabla^2\vc{v}-\nabla p+\nabla\cdot\vc{\sigma}^S_{el}+\frac{1}{2}\eta'\nabla\times\nabla^2\vc{\omega}_p=0,
\end{equation}
where $\vc{\sigma}^S_{el}$ is the symmetric part of the Maxwell stress tensor,
\begin{equation}
\label{2.14}\vc{\sigma}^S_{el}=\frac{1}{2}(\vc{D}\vc{E}+\vc{E}\vc{D})-\frac{\varepsilon_1}{2}E^2\vc{I}.
\end{equation}
Using Maxwell's equations of electrostatics one may express the divergence of this tensor as \cite{14}
\begin{equation}
\label{2.15}\vc{F}=\nabla\cdot\vc{\sigma}^S_{el}=\varepsilon_1\vc{P}\cdot(\nabla\vc{E})+\frac{\varepsilon_1}{2}\nabla\times(\vc{P}\times\vc{E}).
\end{equation}
The first term on the right is the Kelvin force density. The second term may be expressed as the divergence of an antisymmetric tensor. For our purposes the alternative expression \cite{15}
\begin{equation}
\label{2.16}\vc{F}=\frac{\varepsilon_1}{2}\nabla(\vc{P}\cdot\vc{E})-\frac{\varepsilon_1}{2}\vc{E}\times(\nabla\times\vc{P})-\frac{1}{2}\vc{D}(\nabla\cdot\vc{P})
\end{equation}
will also be useful.

The reduced equations of motion (2.12) and (2.13) must be supplemented with boundary conditions for $\vc{v}$ and $\vc{\omega}_p$ at the surface of the sphere. We assume that $\vc{v}$ and $\vc{\omega}_p$ satisfy the no-slip conditions
\begin{equation}
\label{2.17} \vc{v}\big|_{r=a+}=0,\qquad\vc{\omega}_p\big|_{r=a+}=0.
\end{equation}
The field $\vc{E}$ is assumed to vanish for $r\rightarrow\infty$. Together with Maxwell's equations of electrostatics (2.5) and the polarization relaxation equation (2.10) the equations constitute a nonlinear set. We solve the equations by formal perturbation expansion in powers of the amplitude of the exciting charge density $\rho_{el}$, putting
\begin{eqnarray}
\label{2.18}\vc{E}&=&\vc{E}_1+\vc{E}_3+...,\qquad\vc{P}=\vc{P}_1+\vc{P}_3+...,\nonumber\\
\vc{v}&=&\vc{v}_2+\vc{v}_4+...,\qquad p=p_0+p_2+p_4+...,\qquad\vc{\omega}_p=\vc{\omega}_{p2}+\vc{\omega}_{p4}+...,
\end{eqnarray}
where $p_0$ is the static equilibrium pressure, and the subscripts denote the power of $\rho_{el}$. We perform the calculation to second order in $\rho$. The higher order terms $(\vc{E}_3,\vc{P}_3,\vc{v}_4,p_4,\vc{\omega}_{p4},...)$ are generated by the two convective terms in Eq. (2.10) and by nonlinearity in the equation of state $\vc{P}_{eq}(\vc{E})$.

\section{\label{III}First order fields and second order propulsion velocity}

The first order electrical field $\vc{E}_1$ is expressed by Eq. (2.6) in terms of the external multipole moments $\mu_1,\mu_2$. To first order the flow velocity $\vc{v}_1$ and the particle rotational velocity $\vc{\omega}_{p1}$ vanish, so that for the calculation of the first order polarization $\vc{P}_1$ the convective terms in Eq. (2.10) can be omitted. The linear relaxation equation reads
\begin{equation}
\label{3.1}\frac{\partial\vc{P}_1}{\partial t}=-\gamma(\vc{P}_1-\chi_0\vc{E}_1),
\end{equation}
where $\chi_0$ is the zero field susceptibility.
We decompose the fields $\vc{E}_1$ and $\vc{P}_1$ as in Eq. (2.9)
\begin{eqnarray}
\label{3.2}\vc{E}_1(r,\theta,t)&=&\vc{E}_{1c}(r,\theta)\cos\omega t+\vc{E}_{1s}(r,\theta)\sin\omega t,\nonumber\\
\vc{P}_1(r,\theta,t)&=&\vc{P}_{1c}(r,\theta)\cos\omega t+\vc{P}_{1s}(r,\theta)\sin\omega t.
\end{eqnarray}
It is convenient to use complex notation with oscillating factor $\exp(-i\omega t)$. Then with linear susceptibility $\chi=\chi'+i\chi''$ the field and polarization components are related by
\begin{equation}
\label{3.3}\vc{P}_{1c}=\chi'\vc{E}_{1c}-\chi''\vc{E}_{1s},\qquad\vc{P}_{1s}=\chi'\vc{E}_{1s}+\chi''\vc{E}_{1c}.
\end{equation}
We find from Eq. (3.1)
\begin{equation}
\label{3.4}\chi'=\chi_0\frac{\gamma^2}{\omega^2+\gamma^2},\qquad \chi''=\chi_0\frac{\omega\gamma}{\omega^2+\gamma^2}.
\end{equation}
The linear susceptibility is used in the dielectric problem mentioned below Eq. (2.9).

We find for the second order electrical torque density
\begin{equation}
\label{3.5}\vc{N}_2=\varepsilon_1\vc{P}_1\times\vc{E}_1=(0,0,N_{2\varphi}),
\end{equation}
with
\begin{equation}
\label{3.6}N_{2\varphi}=C\frac{a^7}{r^7}(5\sin\theta+\sin3\theta),\qquad C=\frac{3}{8}\varepsilon_1\chi''(\mu_{1c}\mu_{2s}-\mu_{1s}\mu_{2c}),
\end{equation}
independent of time. The torque density (3.5) is the central quantity in our derivation. The torque density acts on the fluid, creating a ring vortex surrounding the sphere and causing it to move. In Fig. 1 we show a schematic picture of the effect. In order to find the propulsion velocity of the sphere we must calculate the vortex flow pattern from the equations of motion for the fluid.

To second order the equations of motion Eqs. (2.12) and (2.13) become
 \begin{eqnarray}
\label{3.7}2\zeta(\nabla\times\vc{v}_2-2\vc{\omega}_{p2})=-N_{2\varphi}\vc{e}_\varphi-\eta'\nabla^2\vc{\omega}_{p2},\nonumber\\
\eta\nabla^2\vc{v}_2-\nabla p_2+\nabla\cdot\vc{\sigma}^S_{el2}+\frac{1}{2}\eta'\nabla\times\nabla^2\vc{\omega}_{p2}=0.
\end{eqnarray}
In the term $\nabla\cdot\vc{\sigma}^S_{el2}$ we can use Eq. (2.16) with $\vc{P}$ and $\vc{E}$ replaced by $\vc{P}_1$ and $\vc{E}_1$. The last two terms in the expression then vanish, because $\nabla\times\vc{P}_1=0$ and $\nabla\cdot\vc{P}_1=0$. The first term shows that the electric force density can be balanced by the gradient of a pressure. Therefore we look for a solution of Eqs. (3.7) with the term involving $\nabla\cdot\vc{\sigma}^S_{el2}$ omitted. The remaining equations are driven by the time-independent torque density $\vc{N}_2$, and we denote the corresponding remaining pressure disturbance as $p_{2N}$.

The geometry of the torque density suggests that the flow velocity is axially symmetric. We can reduce the equations to scalar form by putting $\vc{\omega}_{p2}=(0,0,\chi(r,\theta))$ and using a Stokes stream function $\psi(r,\theta)$ such that $\vc{v}_2=(v_{r},v_{\theta},0)$ with
\begin{equation}
\label{3.8}v_r=\frac{-1}{r^2\sin\theta}\frac{\partial\psi}{\partial\theta},\qquad v_\theta=\frac{1}{r\sin\theta}\frac{\partial\psi}{\partial r}.
\end{equation}
The angular factor in the torque density in Eq. (3.6) can be expressed as
\begin{equation}
\label{3.9}5\sin\theta+\sin3\theta=\frac{24}{5}P^1_1(\theta)+\frac{8}{15}P^1_3(\theta),
\end{equation}
with
\begin{equation}
\label{3.10}P^1_1(\theta)=\sin\theta,\qquad P^1_3(\theta)=\frac{3}{8}(\sin\theta+5\sin 3\theta).
\end{equation}
This suggests that we look for a solution of the form
 \begin{eqnarray}
\label{3.11}\psi(r,\theta)&=&rf_1(r)\sin\theta\; P^1_1(\theta)+rf_3(r)\sin\theta\; P^1_3(\theta),\nonumber\\
\chi(r,\theta)&=&g_1(r)P^1_1(\theta)+g_3(r)P^1_3(\theta),\nonumber\\
p_{2N}(r,\theta)&=&\eta\frac{h_1}{r^2}P_1(\theta)+\eta\frac{h_3}{r^4}P_3(\theta).
\end{eqnarray}
We have used that the pressure disturbance $p_{2N}$ satisfies Laplace's equation.
Substitution of these expressions leads to two pairs of ordinary differential equations for the pairs $(f_1,g_1)$ and $(f_3,g_3)$ separately.
The equations for the pair $(f_1,g_1)$ read
\begin{eqnarray}
\label{3.12}
2\zeta\big[r^2f''_1+2rf'_1-2f_1\big]+\eta'\big[r^2g''_1+2rg'_1-2g_1\big]-4\zeta r^2g_1=-\frac{24}{5}C\frac{a^7}{r^5},\nonumber\\
2\eta\big[r^2f''_1+2rf'_1-2f_1\big]-2\eta h_1-\eta'\big[r^2g''_1+2rg'_1-2g_1\big]=0.
\end{eqnarray}
The equations for the pair $(f_3,g_3)$ read
\begin{eqnarray}
\label{3.13}
2\zeta\big[r^2f''_3+2rf'_3-12f_3\big]+\eta'\big[r^2g''_3+2rg'_3-12g_3\big]-4\zeta r^2g_3=-\frac{8}{15}C\frac{a^7}{r^5},\nonumber\\
2\eta\big[r^2f''_3+2rf'_3-12f_3\big]-\eta\frac{2h_3}{3r^2}-\eta'\big[r^2g''_3+2rg'_3-12g_3\big]=0.
\end{eqnarray}

Consider first the equations for the pair $(f_1,g_1)$. We can find a particular solution of the second equation in Eq. (3.12) by putting $h_{1p}=0$ and $g_{1p}(r)=(2\eta/\eta')f_{1p}(r)$. Substituting this into the first equation we obtain a second order inhomogeneous equation for $f_{1p}(r)$ of the form
\begin{equation}
\label{3.14}r^2f''_{1p}+2rf'_{1p}-2f_{1p}-\kappa^2r^2f_{1p}=-\frac{12}{5(\eta+\zeta)}C\frac{a^7}{r^5},
\end{equation}
with the abbreviation
\begin{equation}
\label{3.15}\kappa^2=\frac{4\eta\zeta}{\eta'(\eta+\zeta)}.
\end{equation}
The equation has the solution
\begin{equation}
\label{3.16}f_{1p}(r)=\frac{12a^7}{5(\eta+\zeta)}\;C\frac{2\kappa}{\pi}\big[i_1(\kappa r)L_1(r)+k_1(\kappa r)G_1(a,r)\big]+A_1k_1(\kappa r),
\end{equation}
with modified Bessel functions \cite{13A}
\begin{equation}
\label{3.17}i_l(z)=\sqrt{\frac{\pi}{2z}}I_{l+\frac{1}{2}}(z),\qquad k_l(z)=\sqrt{\frac{\pi}{2z}}K_{l+\frac{1}{2}}(z),
\end{equation}
integrals
\begin{equation}
\label{3.18}L_1(b)=\int^\infty_b\frac{k_1(\kappa r)}{r^5}\;dr,\qquad G_1(a,b)=\int^b_a\frac{i_1(\kappa r)}{r^5}\;dr,
\end{equation}
and a constant $A_1$. The integrals can be performed explicitly. The constants of integration have been chosen such that $f_{1p}(r)$ tends to zero at infinity.

In order to satisfy the boundary conditions we must add solutions of the homogeneous Eq. (3.12) with right hand side put equal to zero. The solution with proper behavior at infinity takes the form
\begin{equation}
\label{3.19}f_{1}(r)=f_{1p}(r)-\frac{12a^7}{5(\eta+\zeta)}C\bigg[A_2r+\frac{A_3}{r^2}\bigg],\qquad
g_1(r)=\frac{2\eta}{\eta'}f_{1p}(r),\qquad h_1=0.
\end{equation}
The solution proportional to $A_2$ corresponds to a flow pattern with uniform flow velocity and vanishing pressure. The solution proportional to $A_3$ corresponds to a dipolar irrotational flow pattern, again with vanishing pressure. We have omitted an Oseen flow pattern proportional to $h_1$ with $f_1=-h_1/2$ and $g_1=h_1/(2r^2)$ since such a contribution would imply that the sphere exerts a force on the fluid, which is excluded in self-propulsion. The velocity of self-propulsion is given by minus the uniform flow velocity at infinity, and is proportional to the coefficient $A_2$. The sphere is propelled by a ring vortex generated by the azimuthal torque density proportional to $\vc{e}_\varphi$. The constants of integration $A_1,A_2,A_3$ are determined by applying the no-slip boundary conditions Eq. (2.17). These imply
\begin{equation}
\label{3.20}f_1(a+)=0,\qquad f'_1(a+)=0,\qquad g_1(a+)=0.
\end{equation}
In particular we find for the coefficient $A_2$
\begin{equation}
\label{3.21}A_2=\frac{24+24\sigma-6\sigma^2+2\sigma^3-\sigma^4+\sigma^5-\sigma^6e^\sigma\Gamma(0,\sigma)}{432a^6(1+\sigma)},\qquad \sigma=\kappa a.
\end{equation}
The velocity of self-propulsion is given by
\begin{equation}
\label{3.22}\vc{U}_2=U_2\vc{e}_z,
\end{equation}
with scalar
\begin{equation}
\label{3.23}U_2=\frac{-4a}{15(\eta+\zeta)}\;CF(\sigma),\qquad F(\sigma)=18a^6A_2.
\end{equation}
The coefficient $C$ is given in Eq. (3.6) and can be positive or negative.
In Fig. 2 we plot $F(\sigma)$ as a function of $\sigma$. The function has the properties
\begin{equation}
\label{3.24}F(0)=1,\qquad F(\sigma)=\frac{6}{\sigma}+O(\sigma^{-2})\qquad\mathrm{as}\;\sigma\rightarrow\infty,
\end{equation}
showing a slow decay for small spin viscosity $\eta'$. 

With the expression (3.23) for the propulsion velocity we have attained the goal of our calculation. The power required to achieve the propulsion velocity is purely electrical, and can be calculated from the first order polarization $\vc{P}_1$ and the electrical field $\vc{E}_1$, given by Eqs. (3.2) and (3.3). Since the first order flow velocity vanishes, the dissipation due to viscosity does not contribute to the order considered. The power equals the time-averaged dissipation given by
\begin{equation}
\label{3.25}\overline{\mathcal{D}}=\frac{1}{T}\int^T_0\int_{r>a}\varepsilon_1\vc{E}_1\cdot\frac{\partial\vc{P}_1}{\partial t}\;dtd\vc{r},
\end{equation}
with period $T=2\pi/\omega$. Substituting for the first order field and polarization and performing the integrations we find
\begin{equation}
\label{3.26}\overline{\mathcal{D}}=\frac{2\pi}{15}\varepsilon_1\omega a^3\chi''\big[10(\mu_{1c}^2+\mu_{1s}^2)+9(\mu_{2c}^2+\mu_{2s}^2)\big].
\end{equation}
Like the propulsion velocity, the power is proportional to the imaginary part of the susceptibility $\chi''$. For the dimensionless efficiency defined by \cite{18}
\begin{equation}
\label{3.27}E_T=\eta\omega a^2\frac{|U_2|}{\overline{\mathcal{D}}}
\end{equation}
we find
\begin{equation}
\label{3.28}E_T=\frac{3}{4\pi}\frac{\eta}{\eta+\zeta}F(\sigma)\frac{|\mu_{1c}\mu_{2s}-\mu_{1s}\mu_{2c}|}{10(\mu_{1c}^2+\mu_{1s}^2)+9(\mu_{2c}^2+\mu_{2s}^2)}.
\end{equation}
For definiteness we may choose the phase such that $\mu_{1s}=0$. Then the efficiency is maximized for $\mu_{2c}=0$ and $\mu_{2s}=\pm\sqrt{10/9}\mu_{1c}$.

Though the velocity of self-propulsion and the corresponding required power have been determined, we consider for completeness also the solution of Eq. (3.13) corresponding to the higher order angular dependence. One can again find a particular solution of the inhomogeneous equations with $h_{3p}=0$ and $g_{3p}(r)=(2\eta/\eta')f_{3p}(r)$ with
\begin{equation}
\label{3.29}f_{3p}(r)=\frac{4a^7}{15(\eta+\zeta)}\;C\frac{2\kappa}{\pi}\big[i_3(\kappa r)L_3(r)+k_3(\kappa r)G_3(a,r)\big]+A_4k_3(\kappa r),
\end{equation}
integrals
\begin{equation}
\label{3.30}L_3(b)=\int^\infty_b\frac{k_3(\kappa r)}{r^5}\;dr,\qquad G_3(a,b)=\int^b_a\frac{i_3(\kappa r)}{r^5}\;dr,
\end{equation}
and a constant $A_4$. The solution with proper behavior at infinity takes the form
\begin{equation}
\label{3.31}f_{3}(r)=f_{3p}(r)+\frac{A_5}{r^4}-\frac{h_3}{30r^2},\qquad
g_3(r)=\frac{2\eta}{\eta'}f_{3p}(r)+\frac{h_3}{6r^4}.
\end{equation}
The three coefficients $A_4,A_5$ and $h_3$ follow from the three boundary conditions $f_3(a+)=0,f'_3(a+)=0,g_3(a+)=0$, which hold in analogy to Eq. (3.20).

\section{\label{VII}Discussion}

For the known viscosity coefficients of water \cite{16} the screening length $1/\kappa$, defined in Eq. (3.15), equals $2.3\;\mathrm{nm}$. For a planar microbot we estimated for typical values of frequency and length scale a velocity of self-propulsion of the order of several nanometers per second \cite{1}. A similar estimate should be valid for the spherical microbot considered here. It would be of interest to demonstrate the self-propulsion in numerical simulation \cite{17}. For the case of a ferrofluid the previous estimate \cite{1} suggested that experimental realization may be feasible.

We have chosen to discuss only the electrical case in detail. The above estimate shows that experimental realization in a polar liquid like water presumably is not possible. For computer simulations the electrical formulation is to be preferred to the magnetic one. Experimental realization may be attempted for a ferrofluid, where the magnetic formulation applies. The successful experiments on ferrohydrodynamic pumping by Mao and Koser \cite{3},\cite{4} suggest that an attempt may be worthwhile.

As shown above, the analytic solution of the effect in spherical geometry is quite intricate, and has an interest of its own. The coupling of translational and rotational degrees of freedom of the fluid poses a challenging problem. We have limited the calculation to second order perturbation theory, but in principle a fully nonlinear calculation on the basis of self-consistent integral equations, like in the planar case, is possible. For the planar case we found that the second order perturbation theory calculation is quite sufficient from a numerical point of view, and we presume that this is true also in the present case.

\newpage

\section*{Figure captions}

\subsection*{Fig. 1}
Schematics of a sphere being propelled by an electrically or magnetically generated vortex ring We show a cross section in the $xz$ plane and propulsion along the $z$ axis.

\subsection*{Fig. 2}
Plot of the reduced propulsion velocity $F(\sigma)$, defined in Eq. (3.23), as a function of $\sigma=\kappa a$. The parameters $\sigma$ and $\kappa$ are defined in Eqs. (3.15) and (3.21), and $a$ is the radius of the sphere. 

\newpage
\setlength{\unitlength}{1cm}
\begin{figure}
 \includegraphics{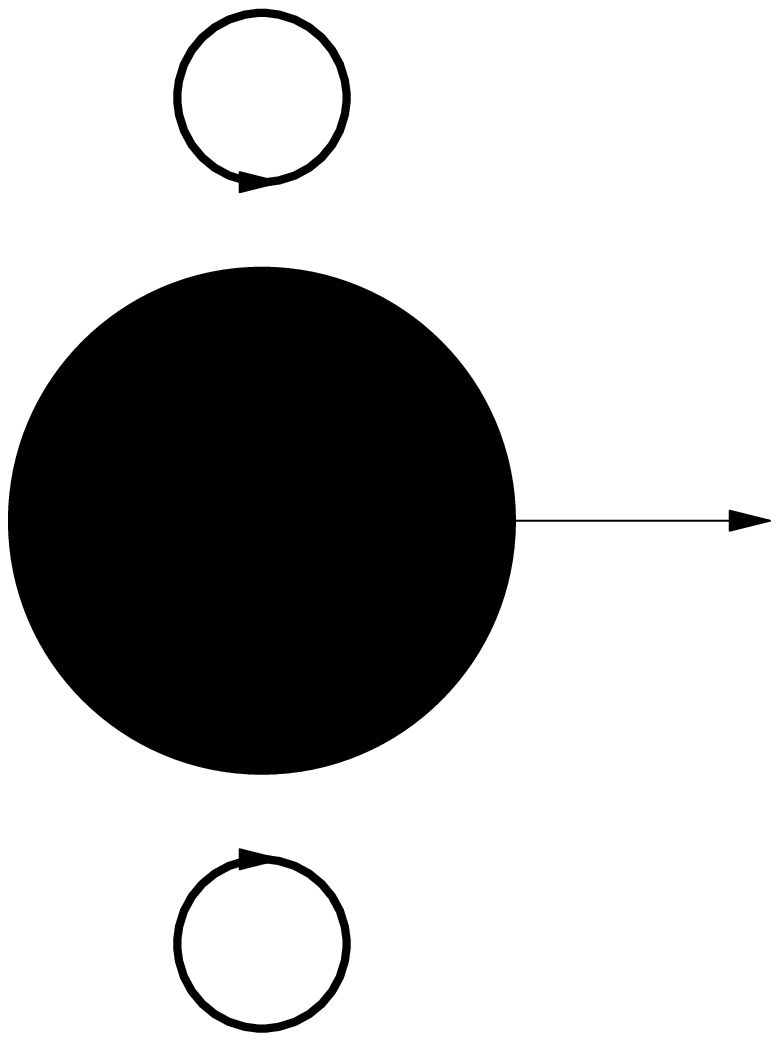}
   \put(-9.1,3.1){}
\put(-1.2,-.2){}
  \caption{}
\end{figure}
\newpage
\clearpage
\newpage
\setlength{\unitlength}{1cm}
\begin{figure}
 \includegraphics{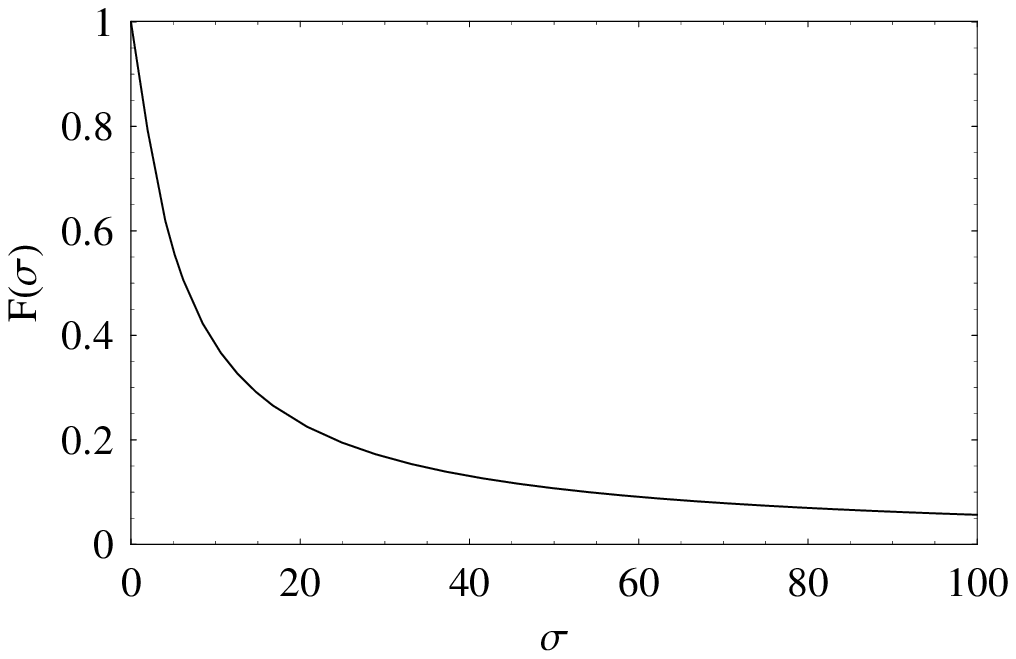}
   \put(-9.1,3.1){}
\put(-1.2,-.2){}
  \caption{}
\end{figure}
\newpage
\clearpage
\end{document}